\documentclass[journal,comsoc]{IEEEtran}
\usepackage{graphicx}
\usepackage{subfigure}
\usepackage{float}
\usepackage{subfloat}
\usepackage{amsmath}
\usepackage{amssymb}
\usepackage{amsthm}
\usepackage{epstopdf}
\usepackage{cases}
\usepackage{color}
\usepackage{makecell}
\usepackage{cite}
\usepackage{booktabs}
\usepackage{diagbox}
\usepackage{multirow}
\usepackage{algorithmic}
\usepackage[ruled]{algorithm2e}

\usepackage[normalem]{ulem}

\usepackage{caption}
\usepackage{mathtools}
\usepackage[T1]{fontenc}
\ifCLASSINFOpdf
\else
\fi

\begin{document}

\title{Beam Training and Tracking in MmWave Communication: A Survey}

\author{Yi~Wang,~\IEEEmembership{Student~Member,~IEEE,}
        Zhiqing~Wei,~\IEEEmembership{Member,~IEEE,}
        and~Zhiyong~Feng,~\IEEEmembership{Senior~Member,~IEEE}
\thanks{This work was supported in part by the National key research and development program under Grant 2020YFA0711302 and the Beijing Natural Science Foundation under Grant L192031.
\emph{(Corresponding authors: Zhiyong Feng, Zhiqing Wei.)}

The authors are with the Key Laboratory of Universal Wireless Communications, Ministry of Education, Beijing University of Posts and Telecommunications, Beijing 100876, China (e-mail: yigg@bupt.edu.cn; weizhiqing@bupt.edu.cn; fengzy@bupt.edu.cn).
}
}
\markboth{}%
{Shell \MakeLowercase{\textit{et al.}}: Bare Demo of IEEEtran.cls for IEEE Communications Society Journals}

\maketitle
\begin{abstract}
Communicating on millimeter wave (mmWave) bands is ushering in a new epoch of mobile communication which provides the availability of 10 Gbps high data rate transmission. However, mmWave links are easily prone to short transmission range communication because of the serious free space path loss and the blockage by obstacles. To overcome these challenges, highly directional beams are exploited to achieve robust links by hybrid beamforming. Accurately aligning the transmitter and receiver beams, i.e. beam training, is vitally important to high data rate transmission. However, it may cause huge overhead which has negative effects on initial access, handover, and tracking. Besides, the mobility patterns of users are complicated and dynamic, which may cause tracking error and large tracking latency. An efficient beam tracking method has a positive effect on sustaining robust links. This article provides an overview of the beam training and tracking technologies on mmWave bands and reveals the insights for future research in the 6th Generation (6G) mobile network. Especially, some open research problems are proposed to realize fast, accurate, and robust beam training and tracking. We hope that this survey provides guidelines for the researchers in the area of mmWave communications.
\end{abstract}

\IEEEpeerreviewmaketitle

\section{Introduction}
\IEEEPARstart{D}{riven} by the emerging bandwidth-intensive applications, such as Ultra-High Definition (UHD) 3D video, Virtual Reality (VR), and Augmented Reality (AR), wireless data traffic is predicted to double every 18 months, which shows an explosive growth trend of ultra-high data rates \cite{ref1,ref2,ref3}. As predicted in \cite{ref4,ref5}, worldwide data traffic demand will grow to 5 Zettabytes (ZBs) per month, with personal data rates reaching 100 Gbps by 2030. How to satisfy such a huge bandwidth requirement for the next generation wireless communication is still a big challenge with limited spectrum resources.

The 6th Generation (6G) era is foreseen to deliver super large-scale connectivity and much higher data rates with more robust reliability than today's legacy systems. As reported in \cite{ref13,ref13-1}, the peak data rate of 6G will reach 1 Tbps, which is 50 times larger than that of the 5th Generation (5G), to support the new applications such as holographic images and autonomous driving. Besides, the latency of 6G will reach 0.1 ms which is one-tenth of 5G. To satisfy such strict requirements, there exist two possible ways to improve the transmission capacity. One is improving the spectrum efficiency by means such as high-order modulation and large-scale Multiple Input Multiple Output (MIMO), which can improve the carrying capacity of the communication system \cite{ref8}. The other is increasing the system bandwidth and continuously enhancing the data service capacity through carrier aggregation, dual connection, and other technologies \cite{ref9}. Even if these technologies have made great efforts to improve the transmission capacity, how to break the bottleneck of wireless bandwidth shortage is a challenge. Millimeter wave (MmWave) bands have rich bandwidth resources to satisfy the requirements of high transmission capacity, promoting the development of wireless communications \cite{ref11}. Due to the diversification of future applications and the intelligence of connections, 6G will generate massive amounts of data for deep processing, which requires huge bandwidth to support \cite{ref14,ref16}. The mmWave bands are considered to be the golden spectrum bands to support 6G \cite{ref18}.

However, there exist several challenges for mmWave communication. \emph{(1) Limited communication range.} The attenuation of mmWave bands in free space is very high due to the high frequencies, resulting in limited effective communication range \cite{ref20}. \emph{(2) Weak diffraction capability.} Owing to the highly directional mmWave link and the sensitivity to be blocked by humans or furniture, it is difficult to propagate through diffraction and other ways, which seriously affects the practical application of mmWave \cite{ref22}. \emph{(3) High training overhead.} As described in the 3rd Generation Partnership Project (3GPP), the array sizes of macro Base Station (BS) are up to 256 elements while the cellular User Equipment (UE) has 32 elements in the new Vehicle-to-Everything (V2X) use cases \cite{ref23,ref24}. A large scale antenna array will cause high training overhead inevitably. \emph{(4) Inconsistent high beamforming gain.} The dominant directions of mmWave often vary fast owing to the mobility and the variations of radio environment. Moreover, the motion patterns of users are usually complicated and dynamic, which may cause tracking error and wasted resources for link establishment and maintenance. Keeping good mmWave communication quality requires to provide high tracking accuracy continuously in various fading scenarios.

In order to overcome these challenges, one possible option is to use a large-scale phased antenna array to create a highly directed beam using beamforming \cite{ref26,ref27}. Notably, researchers had found that the energy of mmWave signal can concentrate into directional narrow beam for data transmission by combining mmWave and beamforming technology, which not only improves the signal transmission range but also realizes the space division multiplexing of spectrum resources and improves spectrum utilization \cite{ref25}. This type of spatial multiplexing technology transforms omnidirectional signal coverage into precise directional service with minimal interference between beams, which is able to serve different communication links in the same space, substantially improving the service of BS. To fully exploit the benefits of directional transmission, the transmitter and receiver beams must be perfectly aligned \cite{ref31,ref32}. Besides, beam training is employed to acquire immediate Channel State Information (CSI) and determine the strongest channel path, so that an acceptable communication quality can be sustained \cite{ref33}. However, the following challenges need to be solved to achieve beam training and tracking. \emph{(1) Swiftness of beam switch.} Owing to the rapid changing of the wireless channel, the beam training of such ``pencil beams'' causes huge overhead, especially in high mobility scenarios \cite{ref35,ref36}. The Angle of Departure (AoD) and Angle of Arrival (AoA) change rapidly which means that the beam training needs to be executed swiftly. \emph{(2) Accuracy of beam selection.} In \cite{ref34}, with a 7-degree beamwidth, an 18-degree misalignment decreases the link budget by about 17 dB, decreasing the maximum throughput by up to 6 Gbps or destroying the connection, according to the experimental data. \emph{(3) Robustness of beam quality maintenance.} Considering high mobility scenario, the best beam will change with the user's location. Thus, the best beam needs to be constantly switched to provide users with seamless coverage and ensure that communication is not interrupted. The temporal correlations of the AoD/AoA can be employed to facilitate beam training by tracking the variation of AoD/AoA. In order to meet future requirement of high data rate, a strong communication link needs to be established efficiently, implying that the process of beam training and tracking needs to be realized swiftly, accurately, and robustly \cite{ref37}.

There exist several survey articles on mmWave technologies in recent years, which contain the aspects of the propagation models in \cite{ref39}, the utilization of mmWave in 5G networks in \cite{ref40-1,ref40-2}, and beam management framework in \cite{ref41}. Rappaport \emph{et al.} provided an in-depth survey on mmWave propagation models \cite{ref39}. \cite{ref40-1} and \cite{ref40-2} explore the exploration of feasibility, advantages, and challenges of mmWave communication in 5G network. For mmWave cellular networks, Giordani \emph{et al.} presented an overview of the measurement collection framework and mobility management \cite{ref41}. Although the feasibility, advantages, and challenges of mmWave communications have been revealed in above mentioned works, none of them concentrate on the concrete technologies about beam training and tracking, which is the first step to employ mmWave communication in future wireless network. The objective of this article is to offer a comprehensive survey on existing beam training and tracking techniques for mmWave communication. Owing to the rapid development of mmWave communication, we incorporate recent work on beam training and tracking into this article to promote the understanding of mmWave beam training and tracking development trends. The key contributions of this article are summarized as follows.

\begin{itemize}
\item The application scenarios are provided to show how mmWave communications leverage its unique features to meet the requirements of services.
\item An in-depth and comprehensive analysis of beam training and tracking technologies in terms of training overhead, tracking complexity, tracking error and advantages and disadvantages is provided.
\item The existing methods of beam training and tracking technologies are summarized, along with some open research problems for further research.
\end{itemize}

The rest of this article is laid out as follows. Section II presents the scenarios that can benefit from mmWave communication and the requirements of beam training and tracking. Section III provides the propagation model of the mmWave signal. Section IV describes the state-of-the-art of beam training in detail. A detailed review of beam tracking algorithms is provided in Section V. Open research problems on the beam training and tracking of mmWave communication are discussed in Section VI. Finally, Section VII concludes this article.

\section{Application Scenarios}
Before discussing the existing research on beam training and tracking technologies, we first list the various application scenarios that have benefited from the beam training and tracking technology currently and in the future, including Wireless Local/Personal Area Network (WLAN/WPAN), cellular network, Vehicle-to-Everything (V2X) communications, High Speed Train (HST), and Unmanned Aerial Vehicle (UAV) etc \cite{sce1}.  As shown in Figure 1, the mmWave network integrates WLAN/WPAN, Cellular, V2X, HST, and UAV scenarios, aiming to offer stable and reliable services for terminals by employing mmWave directional transmission. Enormous data processing request has emerged with newly proposed applications such as ship navigation, positioning remote real-time sensing, cooperative detection, and information fusion, prompting mmWave technologies to be employed in ground-to-ground, air-to-air, and air-to-ground links.

\begin{figure}[!t]
	\centering
	\includegraphics[width=0.35\textheight]{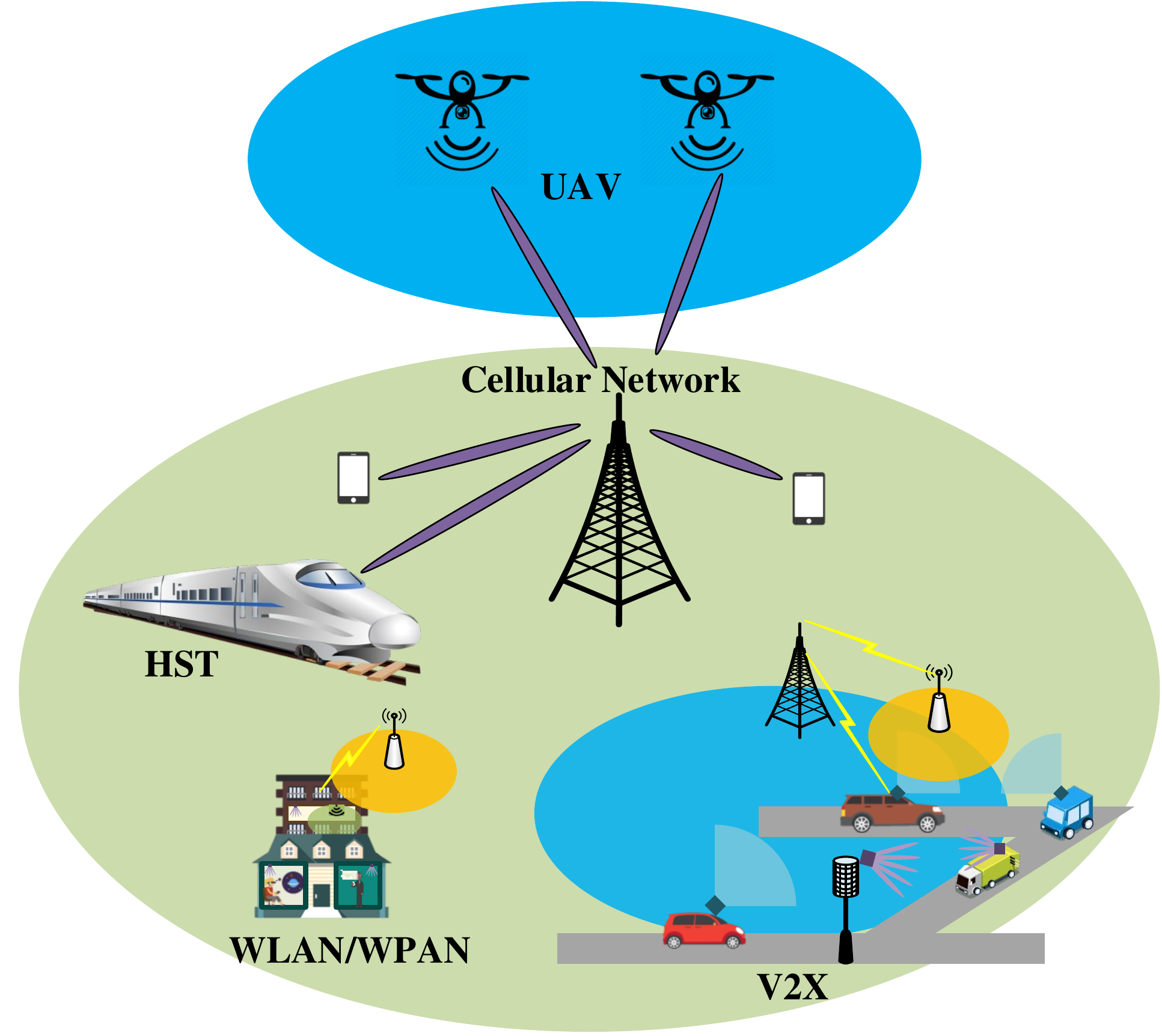}%
	\DeclareGraphicsExtensions.
	\caption{The application scenarios of mmWave communication}
	\label{fig:1}
\end{figure}

\subsection{WLAN/WPAN}
Since the standards of IEEE 802.15.3c and IEEE 802.11ad are developed for WLAN or WPAN communication technologies in the 60 GHz band, mmWave communications have been applied in WLAN or WPAN directly \cite{sce2,sce3}. Broadband multimedia applications data transmission may be realized with the enormous unlicensed bandwidth of 60 GHz, and the transmission rate of Gbps can be reached over short distances with low-cost implementation \cite{sce4}. However, in the 60 GHz band, high path loss up to 15 dB/km not only greatly reduces the mutual interference and improves the security of the link, but also makes the link budget one of the challenging issues \cite{sce4-1}. Besides, The high directivity of beamforming makes the link between the access point and the device extremely fragile. Slight misalignment or artificial obstruction may lead to frequent link interruption, which seriously affects the stability of the wireless link.

\subsection{Cellular Network}
To meet the massive rise in data traffic and connectivity, the 3GPP has completed the standardization process for Release 15 of New Radio (NR) access technology \cite{sce5}. With the continuous increase of mobile communication volume and such a huge connection density of ${10^6} /{\rm{k}}{{\rm{m}}^2}$, the demand for peak data rate even reaches 10 Gb/s, which promotes the continuous increase of wireless channel bandwidth \cite{sce6}. To cope with these challenges, mmWave has the characteristics of short wavelength, narrow beam, flexible and controllable, which is a promising candidate to sustain an acceptable communication quality. Besides, Multi-User MIMO (MU-MIMO) technology enables the base station to transmit data for multiple users simultaneously by taking advantage of the spatial independence of each user which can make full use of space resources, so that the system capacity and transmission rate have been significantly improved \cite{sce7}. Feng \emph{et al.} explore the potential gain of ultra-densification for enhancing mmWave communications from a network-level perspective to realize a sufficient mmWave link margin \cite{sce9}. In addition, high MIMO array gain can not only counteract the increased path loss of mmWave communication, but also is essential to provide an appropriate link budget \cite{sce8}. The current researches reveal a trade-off between the training speed and beamforming performance which can be found in the beam training process.

\subsection{V2X Communications}
With the continuous growth of automobile ownership in recent years, the carrying capacity of road has reached saturation. Traffic safety, travel efficiency, and environmental preservation are all pressing issues that Intelligent Transportation Systems (ITS) may help to tackle. The development of communication technology promotes the development of intelligent transportation systems by realizing V2X communications \cite{scev1}. Based on research in the field of Mobile Ad hoc NETworks (MANET), Vehicular Ad hoc NETworks (VANET) have been the focus of recent research to support V2X communication using Dedicated Short-Range Communication (DSRC) protocol \cite{scev2}. Cellular based V2X (C-V2X), based on cellular communication such as LTE, can provide ubiquitous coverage and support large scale links \cite{scev3}. Some advanced V2X use cases (e.g. vehicle platooning, extended sensor, remote driving, and advanced driving) of NR V2X have put forward 1 Gbps data rate transmission requirements for vehicle communication. For high-level autonomous driving cars, the requirement of delay for collision avoidance is as low as 3 ms, and the transmission rate is required to be more than 1 Gbps, according to 3GPP Release 16 requirements in the ``collective perception of environment'' scenario. Besides, the process of mmWave beam alignment and beam tracking is time-consuming which poses a significant challenge in mobile scenarios \cite{scev6}. During the beam training and tracking process, the position relations among vehicles are obtained from the sensing information which can minimize the beam searching space and decrease the latency of beam training and tracking in contrast to traditional random beam searching algorithms. Besides, the AoD and AoA change significantly in the mobility scenario, leading to the accurate and low-latency beam tracking being a challenging problem. Besides, sensory data may be employed to estimate a vehicle's position and velocity, providing a realistic solution to the requirements of ultra-low end-to-end latency and multi-gigabit-per-second data transfer in the future ITS \cite{scev9}.

\subsection{High Speed Train}
High Speed Train (HST) is anticipated to provide high mobility up to 500 km/h while maintaining adequate Quality of Service (QoS). Actually, HST scenarios have the characteristics of fast movement speed, fast channel change, and large Doppler frequency offset \cite{scet5,scet7}. The combination of mmWave and directional beamforming technology can well meet the needs of high-quality data services in HST scenarios. However, although there exist several researches on the channel model of large-scale MIMO in high-speed mobile scenarios, the measurements, and analytical modeling are mostly based on ideal environments, and further research is needed to model the channels in real HST environments. HST has fast speed and time-varying channels, thus increasing the channel estimation complexity of large-scale MIMO systems and the overhead. Thus, the approaches to decrease channel estimation overhead and computational complexity are required. A fast beam training method is proposed in \cite{scet3} which reduces the beam training subset and compensates the angle offsets caused by movements through fast initial access processes. Moreover, an angle-domain channel tracking method employing linear Kalman filter is proposed to reconstruct CSI, thus compensating the Doppler frequency offset through beam training \cite{scet4}. To guarantee the high throughput and meet the QoS requirement in the HST communications, a dynamic beam tracking scheme is proposed to improve the capacity. Finally, frequent handover problems need to be addressed considering limited coverage of the serving beam and high speed of trains. A joint optimized dynamic inter beam handover method is proposed to optimize the handover failure rate and resource utilization rate \cite{scet6}.

\begin{figure*}[!t]
	\centering
	\includegraphics[width=0.7\textheight]{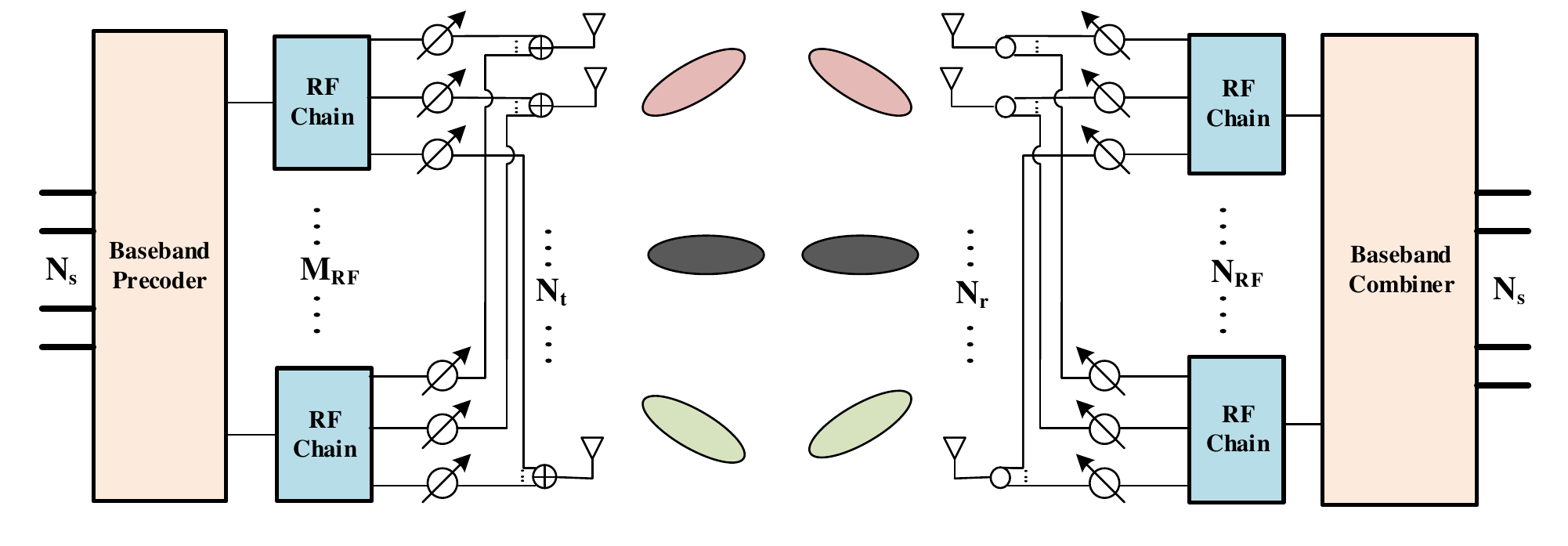}%
	\DeclareGraphicsExtensions.
	\caption{Hybrid MIMO architecture \cite{ref27}}
	\label{fig:2}
\end{figure*}

\subsection{Unmanned Aerial Vehicle}
Due to the characteristics of low cost, rapid deployment and wide coverage, Unmanned Aerial Vehicles (UAVs) are becoming increasingly essential in agricultural plant protection, meteorological and environmental protection monitoring, post-disaster rescue, communication relay, and other civil fields \cite{sceu1}. UAV base station platform can not only collect data (e.g. pictures, videos and files) by hardware devices installed on UAV such as cameras to assist terrestrial users to detect the dangerous situations, but also quickly recover some communication function in emergency circumstances. However, the data rate requirement of services provided to the UAV applications is very large, such as 100 Mbps for 8K video live broadcast and 25 Mbps for Remote UAV controller through HD video \cite{sceu3}. Compared with conventional mmWave beamforming, the beamforming between the UAV and terrestrial users is 3D beamforming. A novel 3D beam training policy to assist UAV mmWave communications is presented to enhance the quality of communication and extend the communication coverage \cite{sceu4}. Compared with terrestrial communications systems, UAVs are more maneuverable in terms of the changes in speed, flight path and flight range, bringing new challenges in beam training and tracking. The higher mobility means the more frequent beam training and tracking process should be performed which increases the overhead of the system. Furthermore, the power allocation of the UAV mmWave communication system needs to be designed to perform beam training, beam tracking, and data transmission owing to the limited power hardware \cite{sceu7}.

\section{MmWave Channel Model}
Owing to the small wavelength at mmWave bands, propagation aspects are special compared with sub-6 GHz bands. It's necessary to understand the channel characteristic of mmWave for further research on beam training and tracking.

Consider the situation of the mmWave propagating in free space. According to Friis' Law \cite{cm1}, the received power is represented as follows.

\begin{equation}
{P_r} = {G_r}{G_t}{(\frac{\lambda }{{4\pi d}})^2}{P_t},
\end{equation}
where ${P_t}$ is the transmit power, $\lambda $ is the wavelength, $d$ is the distance between the transmitter (TX) and the receiver (RX), ${G_t}$ and ${G_r}$ are the transmit and receive antenna gains respectively. Obviously, without the antenna gains that compensated by highly directed large scale antenna arrays, the isotropic path loss of mmWave propagation is significantly greater than low frequencies.

\begin{figure*}[!t]
	\centering
	\includegraphics[width=0.7\textheight]{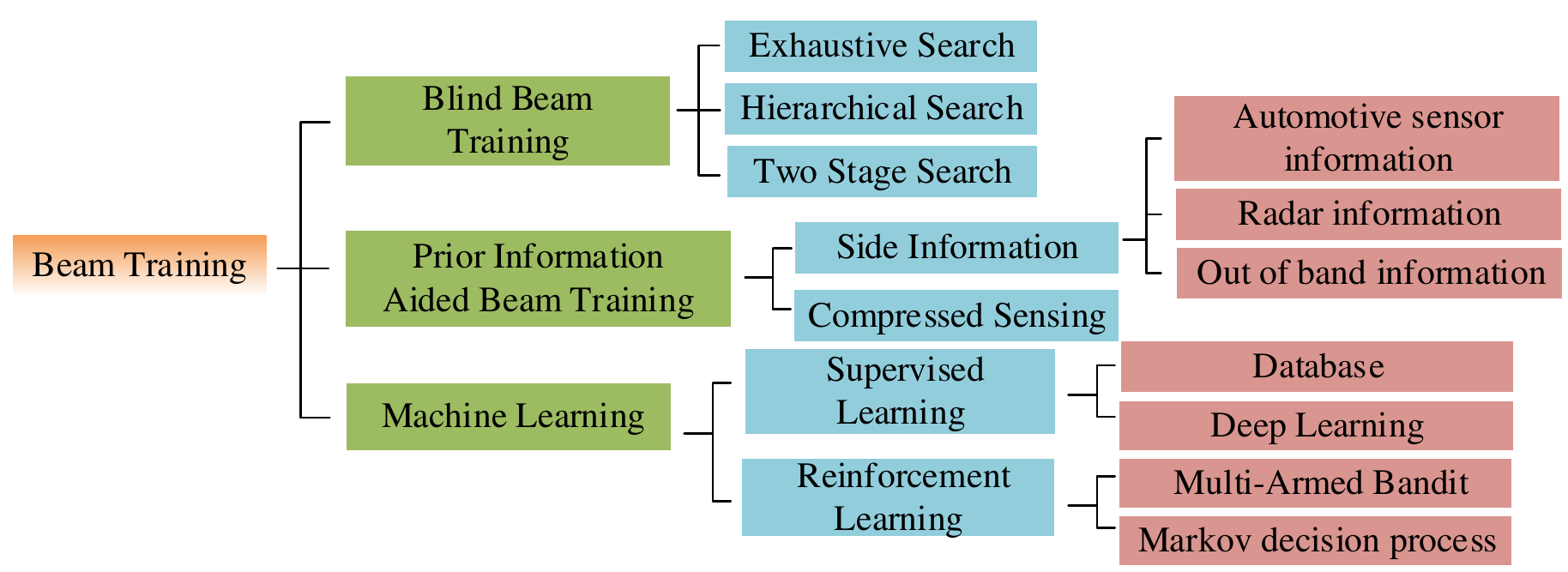}%
	\DeclareGraphicsExtensions.
	\caption{Classification of existing search algorithms for beam training}
	\label{fig:3}
\end{figure*}

The high free space path loss of mmWave transmission leads to limited spatial selectivity and sparse scatterers. Moreover, the antenna correlation is very high due to the tightly arranged large-scale antenna arrays of mmWave transceivers. Combining the above features, a narrow-band cluster model that can accurately capture the structure is proposed, namely Saleh-Valenzuela channel model \cite{ref27,ref32,cm7,cm8}. Consider a mmWave MIMO channel model which can be described by multi-path models. The MIMO system has ${N_t}$ transmit antennas with ${M_{RF}}$ RF chains and ${N_r}$ receive antennas with ${N_{RF}}$ chains. Uniform Linear Arrays (ULAs) are implemented on both the TX and RX in a 2D channel model, with the adjacent antenna spacing $d = \lambda /2$. For an $N$-element ULA, the steering vector is given by
\begin{equation}
a(\theta ) = {\left[ {1,{e^{ - j\frac{{2\pi d\sin (\theta )}}{\lambda }}}, \cdots ,{e^{ - j\frac{{2\pi (N - 1)d\sin (\theta )}}{\lambda }}}} \right]^T}.
\end{equation}

The array phase profile is represented by the array steering vectors ${a_t}({\theta _t})$ and ${a_r}({\theta _r})$, which are related to the angles of departure ${\theta _t}$ and the angles of arrival ${\theta _r}$.

Empolying the clustered channel model, the channel matrix $H$ can be represented as a sum of the contributions of $L$ scattering clusters, which is expressed by a multi-path model \cite{ref27}
\begin{equation}
H = \sqrt {\frac{{{N_t}{N_r}}}{\rho }} \sum\limits_{l = 1}^L {{\beta _l}} {a_r}({\theta _r})a_t^H({\theta _t}),
\end{equation}
where $\rho $ denotes the average path loss, ${{\beta _l}}$ denotes the complex channel gain of the $l$-th path, $L$ denotes the number of propagation paths.

As described in \cite{sceu4}, for a 3D channel model, the steering vector is given by
\begin{equation}
a(\theta ,\varphi ) = {a_{az}}(\theta ) \otimes {a_{el}}(\varphi ),
\end{equation}
where $\theta $ is the azimuth angle, $\varphi $ is the elevation angle and $ \otimes $ is the Kronecker product. The expression of 3D channel model is similar to that of 2D channel model.

For high speed scenarios, i.e., V2X and HST, the high mobility causes a non-negligible Doppler shift. The wireless channel exhibits rapidly time-varying and non-stationary. According to \cite{scet5,scet4,cm9,cm10}, the multipath fading channel is represented as
\begin{equation}
H = \frac{1}{{\sqrt L }}\sum\limits_{l = 1}^L {{g_l}{e^{j2\pi {f_D}t}}} a\left( {{\theta _r}} \right) \odot {b_l},
\end{equation}
where ${g_l}$ denotes the complex small-scale fading gain, ${f_D} = \frac{v}{\lambda }\cos \left( {{\theta _r}} \right)$ denotes the Doppler shift for the $l$-th path, and $v$ denotes the speed of train. $\odot$ denotes the Hadamard product. ${b_l}$ denotes the relative Doppler phase offset vector, which is expressed as
\begin{equation}
{b_l} = {\left[ {1,{e^{j\frac{{2\pi {f_D}d\cos \left( {{\theta _r}} \right)}}{c}}}, \cdots {e^{j\frac{{2\pi ({N_t} - 1){f_D}d\cos \left( {{\theta _r}} \right)}}{c}}}} \right]^T}.
\end{equation}

\section{Beam Training}
Beamforming technology based on MIMO structure has been widely used through which the signals can be transmitted in the form of beam at the TX, so that the transmitted energy can be concentrated in a certain direction to offset the problem of high path loss. A hybrid MIMO architecture is illustrated in Fig. 2. In this configuration, all antenna elements are connected via phase shifters to a
single RF chain, which can support multi-stream transmission with low-cost. At the same time, the RX can also use the beamforming technology to receive signals with high strength in a specific direction. In order to get the direction directed by the receiving and transmitting beamforming, a variety of beam training algorithms are proposed which are described concretely in the following. The number of beam pair alignments during a beamforming setup is defined as the overhead of beam training in this article. Table 1 summarizes the characteristics of several algorithms mentioned below and Table 2 summarizes the key parameters and their notations used in this paper. We summarize the existing beam training methods and divided them into three categories and six subcategories, which is revealed in Fig. 3. Detailed description of existing works are as follows.

\subsection{Blind Beam Training}
It's easy to employ accurate CSI to achieve beam training. However, due to the variability of wireless channel, accurate CSI can't be obtained all the time especially in a low Signal to Interference plus Noise Ratio (SINR) environment. An effective way to realize beam training is blind search, in which no prior information is known. We will introduce three main methods using blind search in detail.

\newcommand{\tabincell}[2]{\begin{tabular}{@{}#1@{}}#2\end{tabular}}
\renewcommand\arraystretch{1.5}
\begin{table*}[!t]
\caption{The characteristics of existing mmWave beam training technologies}
\centering
\label{tab1}
  \begin{tabular}{|c|c|c|c|c|}
  \hline
  Technologies & Scenarios & Reference  & Overhead & Pros \& Cons \\
  \hline
  \multirow{2}{*}{\tabincell{c}{Exhaustive\\ Search}} & WLAN/WPAN & \cite{es1} &  \multirow{2}{*}{${N_t} \cdot {N_r}$} & \multirow{2}{*}{\tabincell{c}{Simple and effective,\\ but with high complexity.}}  \\
  \cline{2-3}
  ~ & Cellular & \cite{es2} & ~ & ~  \\
  \hline

  \multirow{4}{*}{\tabincell{c}{Hierarchical\\ Search}} & \multirow{4}{*}{Cellular} & \multirow{2}{*}{\cite{hs4}} & \multirow{2}{*}{$2\left( {K + {{\log }_2}{N_t}{N_r} - 1} \right)$} & \multirow{4}{*}{\tabincell{c}{Reducing the number of beams\\ need to be measured, but may\\ undergo serious misalignment\\ error propagation.}} \\
  ~ & ~ & ~ & ~&~  \\
  \cline{3-4}
  ~ & ~ & \multirow{2}{*}{\cite{hs8}} & \multirow{2}{*}{{${L_d}({4^{{S_0}}} + 4{\log _2}{N_t} - 4{S_0})$}} & ~ \\
  ~ & ~ & ~ & ~&~ \\
  \hline
  \tabincell{c}{Two Stage\\ Search} & Cellular & \cite{sce11} &  ${N_t} + {N_r}$ & \tabincell{c}{Reducing the number of  beams ,\\ but increase power consumption.} \\
  \hline
  \multirow{3}{*}{\tabincell{c}{Side\\ Information}} & HST & \cite{scet3} &  \multirow{3}{*}{${S_{tr}}$}& \multirow{3}{*}{\tabincell{c}{Limiting the search space in\\ a small area, but need additional\\ sensing equipments.}} \\
  \cline{2-3}
  ~ & V2X & \cite{scev7} & ~ & ~   \\
  \cline{2-3}
  ~ & UAV & \cite{si12} & ~ & ~  \\
  \hline
 \tabincell{c}{ Compressed\\ Sensing} & Cellular & \cite{cs1} &  ${S_{{N_t}}} + {S_{{N_r}}}$ & \tabincell{c}{ Reducing the number of beams\\ need to be measured, but the\\ absence of antenna gain must\\ be addressed during the\\ measurement step.} \\
  \hline
  \multirow{4}{*}{\tabincell{c}{Supervised\\ Learning}} & \multirow{2}{*}{V2X} & \multirow{2}{*}{\cite{sl1}}  & \multirow{4}{*}{${S_{tr}}$} &\multirow{4}{*}{\tabincell{c}{ Wide range of applications and \\high accuracy but with high \\cost to collect training samples. }} \\
 ~ & ~ & ~ & ~&~  \\
  \cline{2-3}
   ~ &   \multirow{2}{*}{Cellular} &  \multirow{2}{*}{\cite{sl2}}  &~ & ~ \\
 ~ & ~ & ~ & ~&~  \\
  \hline
  \tabincell{c}{Reinforcement\\ Learning}  & Cellular & \cite{ref37}  & ${S_{tr}}$ &\tabincell{c}{No training samples are required,\\ but lack of ability of making \\complex decision and utilizing\\ existing domain knowledge.}  \\
\hline
\end{tabular}
\end{table*}

\renewcommand\arraystretch{1.5}
\begin{table}[!t]
\caption{Key Parameters and Notations in Table I}
\centering
\label{tab2}
  \begin{tabular}{|c|c|}
  \hline
  Symbol & Definition \\
  \hline
  $K$ & The number of users \\
  \hline
  ${L_d}$ & The number of MPCs to be estimated \\
  \hline
  ${S_0}$ & The start layer of beam training \\
  \hline
  ${S_{tr}}$ & The promising candidate beam subset \\
  \hline
  ${S_{{N_t}}}$ & The subset of the transmitted beam \\
  \hline
  ${S_{{N_r}}}$ & The subset of the received beam \\
  \hline
  \end{tabular}
\end{table}

\subsubsection{Exhaustive Search}
As the most precise and the simplest beam training method, exhaustive search sweeps all possible combinations one by one between TX and RX, then jointly examine the beam pair corresponding to the dominant path in the predefined beam codebooks \cite{es1,es2}. It is widely used in WLAN/WPAN with simple assumption that the channel remains unchanged during the training time. However, the training overhead scales with the increasing amount of beam pair. For example, for a MIMO system with 64 antennas at the transceiver, 4096 total measurements are needed.

\subsubsection{Hierarchical Search}
Obviously, partially sweeping the beam has the ability to drastically minimize the training overhead. To cover the beam search space, hierarchical beam search uses multi-level codebooks, with fewer wider beams in the lower level and narrower beams in the upper level \cite{hs4,hs5,hs6,hs7,hs8}. TX and RX scan wider beam pairs based on bisection search and find the strongest beam pair, then iteratively reduce the beam search space to the beam subspace related to the strongest beam pair until the precision is achieved. In \cite{hs4}, for the BS, the authors design the hierarchical codebook adaptively for different users rather than using the same one, where the codewords in the current layer are determined by the results of the previous layer. When compared to exhaustive search, hierarchical search has a significantly lower training overhead. However, there is a serious intrinsic error propagation problem with hierarchical search, which implies that if an incorrect beam combination is picked at any predetermined step, all subsequent searches would be useless \cite{hs5,hs6}. In addition, wide beams with small beamforming gain may have a higher probability of causing misalignment in the early stage which can't be applied in low SINR environments directly. In \cite{hs7}, the authors proposed an improved hierarchical codebook to settle the error propagation in transition band between two neighboring beams, in which a threshold is predefined to determine whether continuing to further search in transition band. In \cite{hs8}, the authors designed a novel dynamic hierarchical codebook which can be updated by reducing the contribution of the predicted Multi-Path Components (MPCs) from the original codebook.

\subsubsection{Two Stage Search}
To further improve the high accuracy of beam training, two stage search algorithms for beam training are proposed which use the same beam codebook as exhaustive search to concentrate more energy on possible beam pairs. In \cite{ts1}, the authors proposed an optimized two stage search algorithm to achieve good performance with a limited training budget. In the first stage, the proposed algorithm scans all beam pairs with a small portion of total training energy which is different from conventional equal power allocation. In the second stage, the possible beam combination needs to be measured additionally. The beam combination which has the largest received energy according to the two stage measurement will be the final choice. An adaptive two stage scheme is proposed to improve alignment probability \cite{ts2} in which the beams with poor performance are rejected in the first stage while employing exhaustive search to search the remaining space. In \cite{ts3}, another two stage search is proposed by IEEE 802.11ad standard which decides the possible beam at the user and the BS sequentially by adopting the Sequential Downlink-Uplink (SDU) transmit sector sweep combination. In \cite{sce11}, the authors extend the application scenario to the MU-MIMO case which adopts a Sequential Downlink-Downlink (SDD) transmit sector sweep combination.

\subsection{Prior Information Aided Beam Training}
The exhaustive search always causes huge training overhead which sequentially tests all possible beam pairs. Besides, the dynamic variation of wireless environment brings new challenge to the process of beam training. In the next subsection, fast beam training algorithms have been proposed which utilize prior information of environment such as sensors, DSRC and out of band information to configure the mmWave connection \cite{si1}. The side information aided beam training can reduce the beam search space to the promising directions with environmental awareness. The training overhead can be largely reduced without any performance loss.

\subsubsection{Side Information}
The side information usually contains useful information (e.g., absolute position, velocity) which can be obtained from the sensors or DSRC to reduce the beam search space in a small range. Obviously, there exist three kinds of side information as follows. \emph{(1) Automotive sensor information.} Though the relative position can be known, efficient beam alignment is still needed to avoid the GPS estimation noise or insufficient satellite visibility by adaptively determining proper candidate beams. As described in \cite{si4}, where position information has been considered as a kind of side information, has the ability to assist robust link establishment in mmWave wireless communications. Although there is an inevitable position mistake that causes the uncertainty of AoA/AoD, this may be mitigated by performing a little amount of beam training to cover the uncertainty \cite{si5}. The authors in \cite{scet3} proposed a fast initial access scheme to compensate the angle offset caused by the high mobility of railway where the beam training subset can be updated by the position information together with the latest historical beam training results. \emph{(2) Radar information.} Furthermore, a radar aided beam training is proposed which utilizes the information extracted from the radar echo to configure the mmWave communication beams \cite{si3}. \emph{(3) Out of band information.} Owing to the legacy radios with lower frequencies are expected to be used in association with mmWave systems to improve wide range signaling management and allow multi-band communications, out-of-band information from legacy band has been considered to assist decrease the overhead of establishing the mmWave connections \cite{scev7,si6}. Ali \emph{et al.} and Hashemi \emph{et al.} concurrently exploit legacy band coarse direction finding and compressed channel estimation to enable more accurate beam alignment between each pair of sender and receiver at low cost \cite{si7,si8}. Besides, the promising beam direction can be predicted and updated by the position and the maximum mobility speed to further reduce the overhead of beam training \cite{si9,si12}.

\subsubsection{Compressed Sensing}
Due to the mmWave channel's sparse nature, leveraging compressed sensing to reduce the beam measurements is probably unavoidable \cite{cs1,cs2}. Compressive adaptation based approaches are presented to discover the dominant path (i.e. the strongest multi-path components AoA-AoD pair) and overcome the huge overhead caused by the codebook beam training through leveraging the channel sparsity in both temporal and angular dimensions \cite{cs3}. The AoA/AoD is determined by identifying the position of a non-zero element, which in turn determines the beam training based on compressed sensing. The optimal beam pair with largest beamforming gain can be determined with the accurate CSI. Thus, more and more studies concentrate on the acquisition to assist the determination of beam pointing direction at TX and RX rather than purely blind search or random search \cite{cs4,cs5,cs6}. Conventionally, the path directions can be estimated by MUltiple SIgnal Classification (MUSIC) method. However, because of the existed spectrum ambiguity, the compressed sensing performs better than MUSIC in beam training. In \cite{cs7,cs8}, beam training process is conducted efficiently where wide beams are utilized to search the beam space coarsely while narrower beams are utilized to achieve the desired resolution. The number of measurements for AoA/AoD estimation can be greatly reduced by employing the compressed sensing. In \cite{cs9}, a beam training algorithm is proposed by employing the low rank structure of channel matrix together with the estimated user position which can escape the computation of AoA/AoD compared to convention beam training approaches. In \cite{cs10}, the authors exploited the temporal correlations of the AoA/AoD to figure out that the beam training process is performed frequently in highly dynamic scenarios. In \cite{cs11}, the authors proposed a novel method to identify the AoA/AoD which employs pseudo-random multi-finger beam patterns to search the dominant propagation path. Then the quadratic measurements of the average received power are collected to guarantee the robustness to time variations.

\subsection{Machine Learning Based Beam Training}
Endowing the beam training process with intelligence is a potential technique to decrease the training overhead. Machine learning based beam training can automatically extract and apply relevant information from previous training to limit the search area for subsequent training, rather than explicitly using previous knowledge. Furthermore, due to the superior performance of machine learning in prediction accuracy over conventional technologies, machine learning aided beam training has attracted researchers' attention recently. Generally speaking, there exist two types of machine learning methods, i.e., Supervised Learning (SL) and Reinforcement Learning (RL) methods. The SL algorithms require an offline learning setting, which needs time to collect the training data, causing time consuming especially the data need to be renewed when the environment changes. The RL has the advantage to explore and exploit the information obtained, which is more intelligent compared with SL algorithms.

\subsubsection{Supervised Learning}
Due to the lack of interaction with the environment, SL requires a huge number of training samples to assure its excellent performance \cite{scev7,si6,si9,sl1,sl2,sl3,sl4}. Obviously, there exist two kinds of SL methods as follows. \emph{(1) Database based beam training.} \cite{si9} employed a database containing past beam measurements in specific positions to identify NLOS beamforming directions. A multi-path fingerprint database is formulated in \cite{scev7,si6} for reliable beam training while employing a multi-armed bandit model to make beam selection. In \cite{sl1}, a fingerprint based database is formulated in given positions, then a feedforward neural network is invoked for selection in the database, where the training weights are computed offline. Then, the BS begins the training operation to choose the beam-pair that meets the specified threshold. \emph{(2) Deep learning based beam training.} \cite{sl2} employed a deep learning network to forecast beamforming vectors with the use of signatures carrying the information about surrounding environment. \cite{sl3} proposed a deep learning-based scheme to achieve direction-of-arrival (DoA) estimation and channel estimation by learning the spatial characteristic of the wireless channel. To deal with the non-linear and non-monotonic properties of channel power leakage, two deep neural network based beam training are proposed where the Original DNN-based Beam Training (ODBT) estimates the best beam pair based on the probability vector while the Enhanced DNN-based Beam Training (EDBT) conducts further training after knowing the probability vector \cite{sl4}.

\begin{figure}[!t]
	\centering
	\includegraphics[width=0.35\textheight]{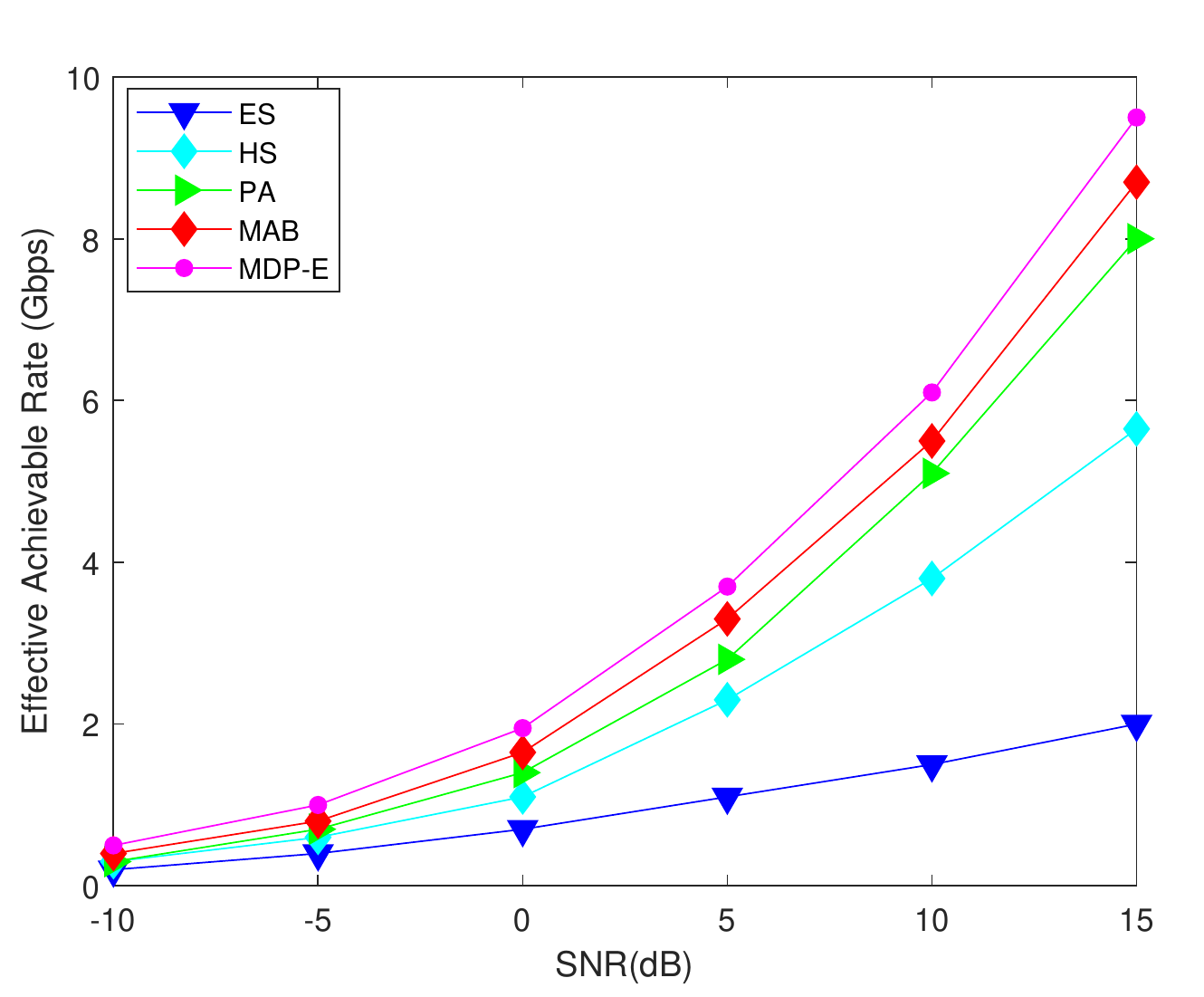}%
	\DeclareGraphicsExtensions.
	\caption{The EAR performance of different beam trainging algoritms}
	\label{fig:4}
\end{figure}

\begin{figure*}[!t]
	\centering
	\includegraphics[width=0.7\textheight]{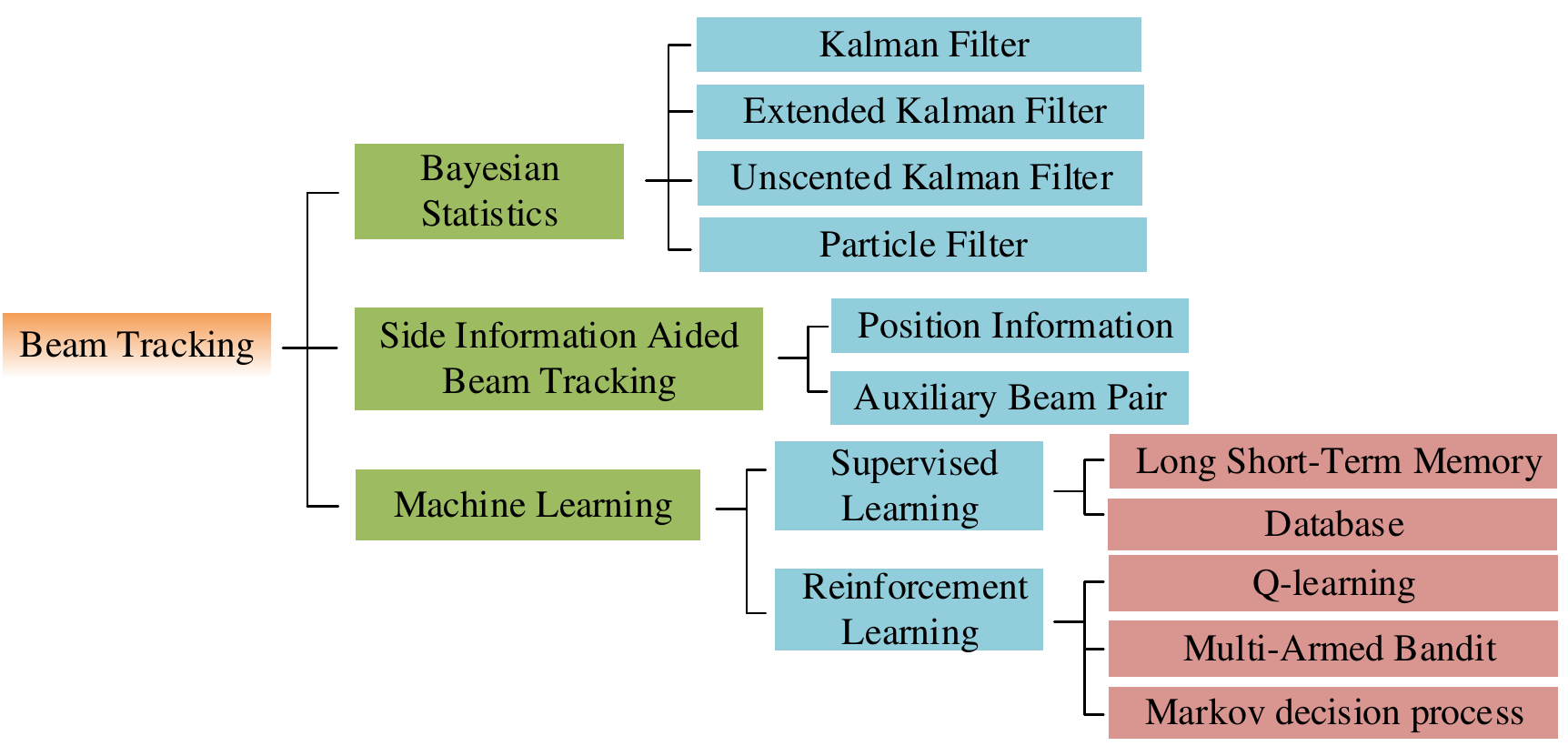}%
	\DeclareGraphicsExtensions.
	\caption{Classification of existing search algorithms for beam tracking}
	\label{fig:5}
\end{figure*}

\subsubsection{Reinforcement Learning}
Obviously, there exist two kinds of RL algorithms as follows. \emph{(1) Multi-Armed Bandit.} More research has been focused on the use of Multi-Armed Bandit (MAB), a lightweight reinforcement learning technique, in beam training problems. A standard MAB issue may be used to simulate the beam training problem. MAB is a useful tool in dealing with the sequential decision-making problems. \cite{rl2} proposed a multiple-play MAB framework which can train multiple beam pairs in each attempt. In the MAB setting, for arms which have high probability of getting high reward, more attempts (exploitation) should be made to obtain a certain cumulative reward. For the unknown arms, a certain amount of exploration must be allocated so as not to miss a higher-yielding option, but at the same time more exploration also means higher risk. This tradeoff is called Exploration-Exploitation (EE) dilemma. Based on a structured MAB model with contextual information, a beam training algorithm can optimally exploit the EE tradeoff. There exists another beam training algorithm which is applied in high speed railways based on an MAB model \cite{rl4}. In \cite{scev8}, a fast machine learning algorithm is proposed by exploiting coarse user location information to model the beam training problem as a contextual MAB problem. \cite{rl5} exploited the contextual stochastic bandits with prior information on the channel fluctuations to reduce the processing time. In \cite{scev7}, different MAB variants are used to define coarse and fine levels in the position-aided beam training process which utilizes a single, common MAB. Besides, a newly risk-aware feature is proposed to reduce the probability of misalignment. A linear Thompson Sampling beam selection policy which combines sparse learning with smart exploration of promising actions is proposed to balance exploration of the promising beam set \cite{rl6}. In \cite{rl7}, a sliding window and a discounting factor are proposed to address the beam training problem in the non-stationary scenario by weighting the recent states higher. In \cite{ref37}, the $\varepsilon $-greedy policy and Upper Confidence Bound (UCB) policy are proposed to keep the balance between EE dilemma. They can adapt to the varying environment without any prior knowledge of dynamic channel modelling. \emph{(2) Markov decision process.} \cite{sce12} combined the adaptive learning ability of machine leaning with the domain knowledge of wireless communications to interactive with the ever-changing environments, making the algorithm adapt to the new environments efficiently.

Figure 4 shows the Effective Achievable Rate (EAR) performance of different beam training algorithms. The Exhaustive Search (ES) based and the Hierarchical Search (HS) based Beam Training algorithms can find the optimal beams but cause huge overhead. The Position-information Aided (PA) beam training algorithm can reduce the search space but has an inevitable position mistake. The MAB based beam training algorithm can't sense the change of the beamforming gain intelligently, which induces the algorithm to take sub-optimal decisions. The Markov Decision Process with Environmental-awareness (MDP-E) based beam training algorithm has the ability to sense the environmental variation and adjust adaptively, which achieves high EAR performance.

\section{Beam Tracking}
Beam training is employed to identify the dominant path to provide optimal beamforming gain. However, owing to the fast-varying environment, frequent beam training will cause huge training overhead. After the directional link is established, slight beam misalignment may cause significant performance loss due to the user mobility, resulting in the reduction of data rates or unexpected link outage \cite{bt1}. Therefore, beam tracking is vital important to maintain the quality of directional communication and help to accelerate the beam training process. However, the efficiency and reliability of beam tracking still need further improvement, especially in a highly dynamic environment. Table 3 summarizes the characteristics of existing beam tracking algorithms in terms of the complexity , tracking error, and pros \& cons of proposed algorithms. The tracking error is represented as the Mean Square Error (MSE) of AoA. Some related parameters and notations are listed in Table 4. We summarize the existing beam tracking methods and divided them into three categories and eight subcategories, which is revealed in Figure 5. Detailed description of existing works are as follows.

\renewcommand\arraystretch{1.5}
\begin{table*}[!t]
\caption{The characteristics of existing mmWave beam tracking technologies}
\centering
\label{tab3}
  \begin{tabular}{|c|c|c|c|c|c|}
  \hline
  Technologies & Scenarios & Reference  & Complexity & \tabincell{c}{Tracking error\\ $[{\rm{ra}}{{\rm{d}}^2}]$} & Pros \& Cons \\
  \hline
  Kalman Filter & Cellular & \cite{kf3} & $\mathcal{O}({N_i}N_s^3)$  & $6 \times {10^{-3}}$ & \tabincell{c}{Small computational volume \\but only for linear systems.}  \\
  \hline
  \tabincell{c}{Extended \\Kalman Filter} &\tabincell{c}{V2X\\ UAV} & \cite{ekf3} &  $\mathcal{O}({N_i}N_s^3+N_s)$ & $2.8 \times {10^{-3}}$ & \tabincell{c}{Wide applicability but without \\considering the uncertainty \\in the linearization.}  \\
  \hline
  \tabincell{c}{Unscented \\Kalman Filter} &\tabincell{c}{V2X\\ UAV} & \cite{ukf6} &  $\mathcal{O}({N_i}N_s^3+N_s)$ & $2.5 \times {10^{-3}}$ & \tabincell{c}{High prediction accuracy but\\ only for Gaussian systems.}  \\
  \hline
 Particle Filter & \tabincell{c}{Cellular\\ V2X\\ UAV} & \cite{pf5} &  $\mathcal{O}({N_p}{N_i}N_s^2)$ & $2.4 \times {10^{-3}}$ & \tabincell{c}{Suitable for non-linear\\ non-Gaussian systems but with\\ high computational volume.}  \\
  \hline
\tabincell{c}{Position\\ Information} &\tabincell{c}{WLAN/WPAN\\ V2X\\ UAV\\ HST} & \cite{pi8} & $\mathcal{O}(\left| D \right|{N_i}N_s^2)$ & $2.6 \times {10^{-3}}$ & \tabincell{c}{Reducing the search space\\ of the beam but needs\\ high accuracy estimation.}  \\
  \hline
\tabincell{c}{Auxiliary \\Beam Pair} &\tabincell{c}{Cellular\\ V2X} & \cite{abp2} &  $\mathcal{O}({N_a}{N_i}N_s^2)$ & $6.5 \times {10^{-3}}$ & \tabincell{c}{High resolution angle \\estimation but with\\ full beam measurements.}\\
  \hline
\tabincell{c}{Supervised\\ Learning} &\tabincell{c}{Cellular\\ V2X\\UAV} & \cite{slk2} &  $\mathcal{O}(4nm+n^2+n)$ & $1.8 \times {10^{-3}}$ & \tabincell{c}{High precision and wide \\applicability but with high\\ cost to collect training samples.}\\
  \hline
\tabincell{c}{Reinforcement\\ Learning} &\tabincell{c}{Cellular\\ V2X\\UAV} & \cite{rlk14} &  $\mathcal{O}({N_i}{\left( {{N_t}{N_r}} \right)^2})$ & $1.5 \times {10^{-3}}$ & \tabincell{c}{Without training large data\\ sets but with weak \\environmental adaptivity.}\\
  \hline
  \end{tabular}
\end{table*}

\renewcommand\arraystretch{1.5}
\begin{table}[!t]
\centering
\caption{Related Parameters and Notations in Table III}
\label{tab2}
  \begin{tabular}{|c|c|}
  \hline
  Symbol & Definition \\
  \hline
  ${N_i}$ & The number of iteration \\
  \hline
  ${N_s}$ & The number of action space \\
  \hline
  ${N_p}$ & The number of sample points \\
  \hline
  $D$ & The training dataset  \\
  \hline
  ${N_a}$ & The number of auxiliary beam pair \\
  \hline
  $m$ & The size of the input array \\
 \hline
  $n$ & The size of the hidden array \\
  \hline
  \end{tabular}
\end{table}
\subsection{Bayesian Statistics Based Beam Tracking}
When tracking users with high mobility, the current position can be predicted from the previous position and velocity. Kalman Filter based beam tracking shows excellent performance in linear Gaussian system and its variants can work well in non-Gaussian system.

\subsubsection{Kalman Filter}
When target tracking is carried out, prior information can be obtained by some positioning technology (such as LiDAR, radar and camera) or the current predicted position according to the position and speed of the previous moment. However, there always exists position error between the real position and the ideal position. To achieve a more precise positioning trajectory, the Kalman Filter (KF) can produce a weighted average of the observation and prediction results as the positioning result. KF is a method that estimates the system state from system input and output data using the state equation of a linear system efficiently. It can be proved that when the tracking problems are linear Gauss process, KF based method is optimal \cite{cs2,kf2}. In \cite{kf3}, a beam tracking method based on KF is suggested to track fluctuations in AoD/AoA and adjust for abrupt changes (e.g., blockage) in propagation pathways. The KF may be used to track not only the line-of-sight (LOS) travel, but also the NLOS path to eliminate their harmful effects. However, the number of measurements required for estimate grows with the size of antenna arrays at the transceiver, according to the KF-based tracking technique. In \cite{rl7}, the authors proposed to employ KF to track the sparse channel support and channel coefficients jointly.

\subsubsection{Extended Kalman Filter}
Extended Kalman Filter (EKF) is the extended vision of KF by employing the first-order linearization of the nonlinear state to solve the tracking problem in nonlinear system. \cite{ekf1} proposed an EKF based beam tracking method to deal with the tracking of time-varying propagation where only one beam pair is performed. In \cite{ekf2}, to further improve the performance of beam tracking, Jayaprakasam \emph{et al.} proposed a robust tracking method based on joint beamforming and EKF. However, the estimation of direction is not accurate by EKF because of its linear approximation. In \cite{ekf3}, Xin \emph{et al.} proposed a robust EKF tracking algorithm by adopting the Minimum Mean Square Error (MMSE) criterion to remove the accumulate error and improve the accuracy of beam tracking. EKF based tracking method can't be applied to current mmWave vehicular communications for its state evolution model is linear and don't consider the influence of kinematic characteristics of vehicles. In \cite{ekf4}, Shaham \emph{et al.} suggested an EKF based tracking system that considers the real world characteristic of vehicles to improve the tracking performance with low computational complexity. Besides, when the linearization of the system is weakened, EKF performs poorly with high computational complexcity.

\subsubsection{Unscented Kalman Filter}
To further ensure the accuracy of the prediction, the Unscented Kalman Filter (UKF) employs the common Kalman filter framework, which can overcome the lack of EKF and provide better performance in linear Gaussian systems. The key of the UKF is that the unscented transform improves the accuracy by nonlinear propagation mean value and covariance. Inspired by the excellent performance of UKF in tracking fast changing AoA/AoD, Liu \emph{et al.} proposed a robust UKF algorithm to estimate AoA/AoD \cite{ukf2}. In \cite{ukf4}, Zhao \emph{et al.} presented a UKF-based tracking approach for tracking the UAV movement status information while simply considering tracking the transmitting beam direction. In \cite{ukf5}, Ge \emph{et al.} employed UKF to track and predict the transmitting and receiving beam direction simultaneously considering the nonlinearity of beamforming. In \cite{ukf6}, Larew \emph{et al.} formulated the state variations of mmWave channels as a linear Gauss-Markov process to make the predictions more accurate.

\subsubsection{Particle Filter}
The Particle Filter (PF) concept is on the basis of Monte Carlo method, which uses a particle set to express probability and may be applied to any state space model. Its core idea is to use a sequential significance sampling technique to explain the distribution of random state particles obtained from a posteriori probability. Owing to the superiority of PF in nonlinear and non-Gaussian systems, it can be widely used for beam tracking under highly mobile scenarios \cite{pf3}. In \cite{pf4}, a couple of tracking methods based on PF is proposed to establish robust and accurate tracking. One is sequential importance resampling PF where a prior function is employed to estimate a particle's likelihood function. The other is auxiliary PF. In order to make the estimation of particles close to reality, an auxiliary variable is added to the estimation of the proposal density. In \cite{pf5}, to prevent tracking failure, a PF-based tracking approach is suggested, in which the beam width is regulated by partial antenna array activation. In \cite{pf6}, a PF based tracking method which employs codebook design and a blind motion prediction to avoid significant training overhead caused by beam sweeping.

\subsection{Side Information Aided Beam Tracking}
The side information can be adopted to reconfigure the directional antenna and switch to the most promising direction, reducing the search space of the beam and avoiding frequent link interruption.

\subsubsection{Position Information}
Compared with exhaustive search based beam tracking scheme, the algorithms with position information can confine the search space in a limited region which can reduce overhead and shorten the training time for link maintenance. Several studies have proposed a solution to beam tracking with sensors to predict the position. In \cite{pi1}, the authors investigated the motion detection in blockage state in mmWave networks with the measurements of sensors. In \cite{pi2}, the authors predicted the beam section and recognized the motion pattern with sensors. In \cite{pi3}, the authors presented a sensor-aided tracking approach that uses solely simulated sensor data. Clearly, fast beam tracking is difficult, especially when using a low sample rate. In \cite{pi4}, the authors proposed a beam updating method with sensor aided where the translation and rotation are decoupled to improve the tracking accuracy. To improve the accuracy of position estimation, a position-aided channel tracking technique is presented, which takes use of the sparse channel characteristic and statistical information of the channel parameters to avoid precoder and combiner reconfiguration \cite{pi5,pi7}. Ke \emph{et al.} proposed a beam tracking approach on the basis of an efficient Gaussian process to predict position information where the search region is confined in the specific area around the predicted position to shorten the tracking time \cite{pi8}.

\subsubsection{Auxiliary Beam Pair}
Many high resolution angle estimation algorithms, such as MUSIC, Estimating Signal Parameter via Rotational Invariance Techniques (ESPRIT), and variations, perform well, but they require a large number of samples to acquire the received signal covariance matrix \cite{abp1}. Recently, high resolution estimation of AoA/AoD can be conducted by pairs of beams \cite{ukf2}. The direction of the pair of beams can be established by comparing their respective amplitudes. In \cite{abp2}, a novel Auxiliary Beam Pair (ABP) based algorithm is proposed to make angle estimation, where employing the two largest received signal strength beams to enhance angle estimation accuracy in high SINR region. An extended work of \cite{abp2} is considering the application of ABP in low SINR region, where employing EKF and ABP to improve the beam tracking performance. Besides, ABP is a model free tracking algorithm which doesn't rely on the angle variation model. ABP can be combined with Q-learning to further reduce the search space \cite{abp5}.

\subsection{Machine Learning Based Beam Tracking}
Beam tracking algorithms based on KF needs to model dynamic channel first. Besides, existing high resolution angle tracking algorithms such as ABP have large complexity due to large search space. The main advantages of Machine Learning (ML) are two folds: \emph{(1)Environment adaptability:} The channel and environment dynamic can be captured well based on the adaptive learning ability of ML. \emph{(2) Low complexity:} The powerful mathematical operations ability of ML reduces the complexity of algorithm.

\subsubsection{Supervised Learning}
To satisfy the increasing demand for communication and strict time delay restrictions, especially with time-varying and unreliable channels, continuous beam tracking is necessary. Supervised Learning (SL) based beam tracking methods are studied to achieve robust and low latency communications for its immunity to nonstationary \cite{pf3}. Obviously, there exist two kinds of SL algorithms as follows. \emph{(1) Long Short-Term Memory.} In \cite{slk1}, the authors adopted a Long Short-Term Memory (LSTM) framework to create the prediction model, which uses historical CSI to estimate and monitor the channel in a vehicle situation. Besides, the authors proposed to employ a LSTM network to analyze the temporal structure and patterns underlying in the time-varying channels and the signals acquired by inertial sensors \cite{slk1-1}. Deep learning-based tracking approaches, in particular, show superior performance in the aforementioned research. In \cite{slk2}, the authors proposed a Deep Neural Network (DNN) based scheme to know the hidden relation between the received training signal and the mmWave channel, and employ LSTM to track the channel. By utilizing LSTM structure, \cite{slk4} predicted the optimal beam by history measurements including SINR or received signal power. \emph{(2) Database.} In \cite{slk5}, Deng \emph{et al.} suggested a beam tracking approach on the basis of a channel fingerprint database, which contains mappings between statistical beamforming gains and users' locations, to improve the beam tracking efficiency.

\subsubsection{Reinforcement Learning}
SL is usually one-time, short-sighted and considers immediate rewards, while Reinforcement Learning (RL) is sequential, long-term and considers cumulative rewards over time. RL allows for more flexible and efficient beam tracking by adapting to the user's behavior and surroundings. There exist three kinds of RL algorithms as follows. \emph{(1) Q-learning.} An online beam tracking scheme proposed by Jeong \emph{et al.} employed deep Q-learning to maximize the received SINR for LOS channels \cite{rlk1}. Besides, owing to the frequent blockage caused by obstacles, NLOS channel must be considered for beam tracking. Wang \emph{et al.} presented an RL based beam tracking algorithm for determining the dominant path at all feasible positions \cite{rlk2}. A model-free Q learning method may be used to solve the beam tracking issue using realistic measurements and acquired rewards, significantly reducing the complexity of tracking \cite{rlk3}. However, it didn't consider simultaneous actions in multiple users scenarios. \emph{(2) Multi-Armed Bandit.} A Combinatorial Multi-Armed Bandit framework is proposed to integrate multiple arms into a super arm to handle the high mobility beam tracking problem \cite{rlk5}. Besides, most existing works have proved that exploiting the temporal correlation of channels can further enhance the beam training efficiency. \emph{(3) Markov Decision Process.} In \cite{cs2}, Duan \emph{et al.} describe the AoA/AoD deviations as tiny Gaussian random variables, assuming that the values of AoA/AoD vary gradually. However, this model cannot account for the sudden changes in AoA/AoD caused by obstacles. Then, according to previous studies, beam tracking can monitor the AoA/AoD change by modeling it as a discrete Markov process. In \cite{cs3}, the reconstruction of the beamspace channel matrix by locating channel supports is described as the AoA/AoD tracking issue, which may be solved using an approximation message passing method. In \cite{rlk7}, the beam tracking protocol based on the channel sparsity characteristic is described as a Partially Observable Markov Decision Process (POMDP), and the training beam sequence is designed. In \cite{rlk9}, the averaged Cramer-Rao Lower Bound (CRLB) is the selection condition for the training beamforming vectors. Furthermore, using the Maximum Likelihood (ML) and Maximum A Posteriori (MAP) criteria, the real direction of AoD is estimated. Moreover, existing work such as \cite{rlk10} employs a single RF chain for beam tracking, while several RF chains are equipped in the mmWave transceivers to catch the rapid variation of AoA/AoD. To combat the uncertainty caused by high mobility in AoD which may lead to a sharp communication performance loss, Zhang \emph{et al.} proposed a codebook beam tracking algorithm that employs wide beams generated by multiple RFs \cite{rlk11}. To improve tracking robustness, a Bayesian based beamformer and tracking technique using an expectation-maximization algorithm is presented in \cite{rlk13}. The estimation of AoD is formulated as a weighted sum of previously observed AoD values and can be near to the real AoD with high probability. In contrast to the above work, Zhang \emph{et al.} provided a broad framework for improving the received signal strength in certain directions by using several TX-RX beam pairs \cite{rlk14}.

\begin{figure}[!t]
	\centering
	\includegraphics[width=0.35\textheight]{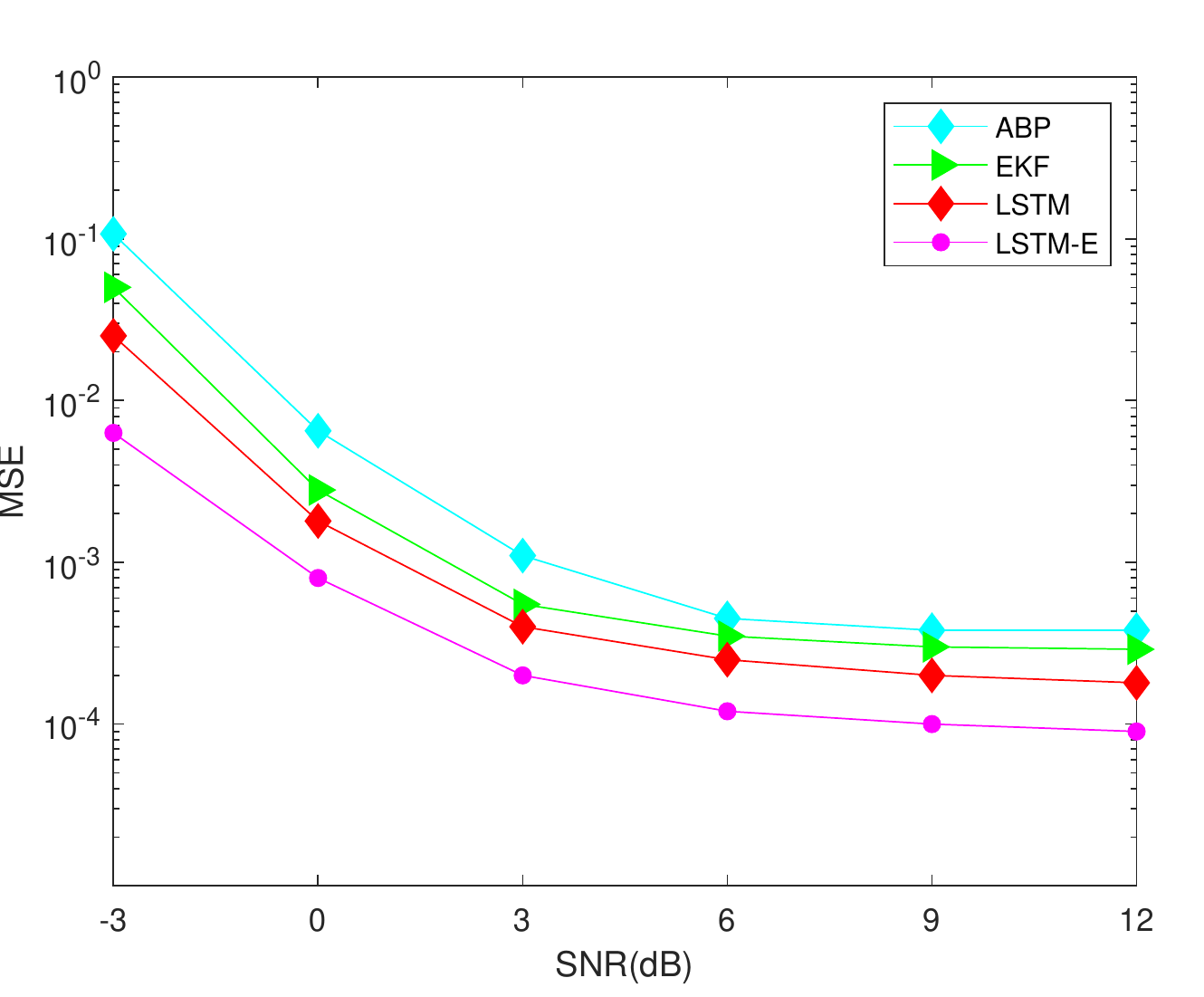}%
	\DeclareGraphicsExtensions.
	\caption{Normalized MSE performance of several beam tracking algorithms}
	\label{fig:6}
\end{figure}

Fig. 6 shows the Normalized MSE of different beam tracking algorithms. The EKF based and the ABP based beam tracking algorithms can achieve a significant gain but are limited to their characteristic that can't capture the behavior of time-varying AoA/AoD sufficiently. The LSTM algorithm performs well in high mobility scenarios. The LSTM with Environmental-awareness (LSTM-E) algorithm can capture the temporal behavior of AoA/AoD caused by the motion of the device, which can track rapidly varying AoA/AoD better.

\section{Open Research Problems}
\subsection{Joint Communication and Sensing}
Joint Communication and Sensing (JCS) system can realize radar detection and communication transmission via shared hardware equipment, which aims to be integrated, miniaturized and utilize spectrum effectively while improving overall performance compared with traditional individual radar and communication devices. In \cite{si3}, the authors used a radar-assisted beam training technique in a V2I scenario, where the radar signal can be used to estimate the covariance of the communication signal, assuming that the primary directions of arrival for the radar and the communication signal are comparable even across bands. In the highly mobile vehicle network, the channel changes rapidly and the cost of channel estimation and transmission is high, feedback based beam training will cause huge overhead \cite{scev8}. Besides, individual radar and communication system not only wastes the spectrum resource, but also needs feedback to confirm whether the proposed beam training algorithm is efficient. Thus, Liu \emph{et al.} proposed a Dual-Functional Radar-Communication (DFRC) system to perform beam tracking \cite{fd1}. Compared with tradition communication-only or radar-only technique, the JCS system has the following advantages.
\begin{itemize}
\item No dedicated downlink pilots are needed.
\item No uplink feedback is needed.
\item Reducing continuous angle estimation error while keeping significant matched-filtering gain.
\item Improving spectrum efficiency and overall system performance.
\end{itemize}

Early contributions for JCS mainly concentrate on the design of integrated signal of JCS system. However, as data traffic grows at an exponential rate and electromagnetic surroundings get more complicated, the shared resources in the JCS system, such as time, spectrum, beam, power and computation, face great challenges in resource scheduling to better serve the target detection, tracking and communication link construction while establishing a good balance between sensing and communication capabilities.

\subsection{Machine Learning}
With abundant spectrum resources, mmWave can achieve Gbps data rate by equipping with large antenna arrays. However, owing to the fast varying channels, beam training and tracking with large number of antennas will cause unaccepted overhead. In contrast to traditional beam training and tracking, the goal of machine learning is to imbue algorithms with intelligence so that relevant data may be collected and utilized automatically, decreasing the beam search space. Machine learning may be used to enjoy its tremendous capacity while simultaneously acquiring situational awareness in order to accomplish quick construction of mmWave connections with low latency and high efficiency \cite{sce12}. Actually, deep learning has great potential in obtaining situational awareness. This includes learning how to capture channel responses, and identifying underutilized spectrum. In addition, deep learning can also be applied to the tasks such as target classification, automatic selection of optimal antennas and RF chains. However, the computational complexity of the beam weights is still high. On the other hand, reinforcement learning is beneficial to solve sequential decision-making problems. Combining deep learning and reinforcement learning technologies will allow the beam training and tracking technologies to be smart, agile, and able to adapt to the fast-varying environment efficiently.

\subsection{THz Beam Training and Tracking}
With the exponential growth of data traffic in wireless communication systems, the terahertz (THz) band, which ranges from 0.1 to 10 THz, is regarded as a viable contender for supporting the data rates of 10 Gbps or possibly several Tbps beyond the 5G networks, spanning the gap between mmWave and optical bands \cite{fd2}. Beamforming is critical for THz communications to correct for signal fading and distortion caused by space loss and multi-path effect during wireless transmission, as well as to decrease interference between users on the same channel. The challenge of THz beamforming design is quite similar to the problem of finding the best beam pair in a typical mmWave system. Though the hybrid architecture of precoding/combining lowers the cost by employing limited RF chains and transfers some signal processing task from the digital baseband signal processor to the analog precoder/combiner, it introduces additional constraints which may weaken the beamforming ability \cite{fd3}. Besides, with the application of massive antenna elements in the future, the characteristics of the wireless channel will change in the spatial domain, namely, the spatial non-stationarity. Thus accurate and practical channel modeling is another challenge for the THz communications. Since beamforming relies on the CSI, the information of AoA/AoD is vital important. Actually, owing to the higher frequencies of THz compared with mmWave frequencies and massive antenna elements, the beam training and tracking endures high computational complexity. How to achieve efficient, accurate, and robust beam training and tracking is urgent to be solved.

\section{Conclusion}
Communicating on mmWave bands breaks the spectrum gridlock and scales up the system capacity for future cellular networks. To achieve rapid, accurate and robust link of mmWave communication, we present an overview of beam training and tracking. Firstly, we list the application scenarios of the mmWave communication. Secondly, we summarize the existing beam training and tracking methods. Finally, we present some open research problems that will be considered in the near future. We hope that this survey provides guidelines for the researchers in the area of mmWave communications.



\ifCLASSOPTIONcaptionsoff
  \newpage
\fi

\end{document}